\documentclass[twocolumn,prl,superscriptaddress]{revtex4-1}

\usepackage{soul}

\usepackage{setspace, graphicx, color, amssymb, amsmath, subfigure, verbatim, ulem}
\usepackage[all]{xy}
\pdfoutput=1

\ifx\pdfoutput\undefined
	\usepackage[hypertex]{hyperref}
\else
	\usepackage[pdftex, hypertexnames=false]{hyperref}
\fi

    \def \T {{\mathbb{T}}}
    \def \e {{\varepsilon}}
    
    \def \g {{\gamma}}
    \def \G {{\Gamma}}
    \def \o {{\theta}}

    \def \k {{\kappa}}

    \def \O {{\mathcal{O}}}
    \def \D {{\Delta}}
    \def \d {{\delta}}

    \def \l {{\lambda}}

    \def \k {{\kappa}}
    \def \n {{\eta}}

\newcommand{\beq}{\begin{equation}}
\newcommand{\eeq}{\end{equation}}
\newcommand{\beqr}{\begin{eqnarray}}
\newcommand{\eeqr}{\end{eqnarray}}
\newcommand{\beqrn}{\begin{eqnarray*}}
\newcommand{\eeqrn}{\end{eqnarray*}}
\newcommand{\beqn}{\begin{equation*}}
\newcommand{\eeqn}{\end{equation*}}
\newcommand{\bei}{\begin{itemize}}
\newcommand{\beii}{\begin{itemize} \item}
\newcommand{\eei}{\end{itemize}}
\newcommand{\bmei}{\begin{itemize} \compactlist}
\newcommand{\emei}{\end{itemize}}
\newcommand{\ben}{\begin{enumerate}}
\newcommand{\een}{\end{enumerate}}
\newcommand{\bes}{\begin{small}}
\newcommand{\ees}{\end{small}}
\newcommand{\bec}{\begin{center}}
\newcommand{\eec}{\end{center}}

\begin{document}

\begin{abstract}
Large networks of sparsely coupled, excitatory and inhibitory cells occur throughout the brain.  A striking feature of these networks is that they are chaotic. How does this chaos manifest in the neural code? Specifically, how variable are the spike patterns that such a network produces in response to an input signal? 

To answer this, we derive a bound for the entropy of multi-cell spike pattern distributions in large recurrent networks of spiking neurons responding to fluctuating inputs. The analysis is based on results from random dynamical systems theory and is complimented by detailed numerical simulations. We find that the spike pattern entropy is an order of magnitude lower than what would be extrapolated from single cells.  This holds despite the fact that network coupling becomes vanishingly sparse as network size grows -- a phenomenon that depends on ``extensive chaos," as previously discovered for balanced networks without stimulus drive. Moreover, we show how spike pattern entropy is controlled by temporal features of the inputs. Our findings provide insight into how neural networks may encode stimuli in the presence of inherently chaotic dynamics.
\end{abstract}

\title{Structured chaos shapes spike-response noise entropy in balanced neural networks}
\author{Guillaume Lajoie$^1$, Jean-Philippe Thivierge$^2$, Eric Shea-Brown$^{3,}$}
\affiliation{University of Washington, Dept. of Applied Mathematics ; $^2$ University of Ottawa, Dept. of Psychology; $^3$ University of Washington, Dept. of Physiology and Biophysics}
\date{\today}
\maketitle


If a time-dependent signal is presented to a network whose dynamics are chaotic and whose initial conditions cannot be perfectly controlled, how much variability can one expect in its responses? Such a scenario is central to questions of stimulus encoding in the brain.  

In this article, we study population level spiking responses in neural networks with sparse, random connectivity and {\it balanced} excitation and inhibition.  Such  networks are ubiquitous models in neuroscience, and reproduce the irregular firing that typifies cortical activity.  Moreover the autonomous activity of such networks is known to be chaotic, with extremely strong sensitivity of spike outputs to tiny changes in a network's initial conditions \cite{Vreeswijk:1998p14451,London:2010p10818,Sun:2010p17792}.  Remarkably, in these autonomous systems, the chaos is invariant to the network scale (i.e., it is {\it extensive}):  the same spectrum of Lyapunov exponents recurs regardless of network size, even when coupling remains localized~\cite{Monteforte:2010p11768,Luccioli:2012p17675}.  Our goal is to add a stimulus drive, and understand the implications for the network spike patterns that result --- a task made challenging by the fact that spikes are related to phase space dynamics in a highly nonlinear way.

Intriguingly, when such chaotic networks respond to time-dependent signals, they produce spiking that is less variable than one might expect (c.f.~\cite{Rajan:2010p7924,PhysRevLett.69.3717}).  In theoretical work, this has been attributed to low-dimensional chaotic attractors that ``project" only intermittently to produce variable spiking in any given single cell~\cite{Lajoie:2013p17297}.  Similar phenomena occur in {\it in vivo} experiments, where fluctuating sensory stimuli are repeatedly presented to an animal.  Here, cortical neurons produce spikes with a wide range of variability, with some spikes repeatedly evoked with millisecond precision~\cite{Yang:2008p17479,Reinagel:2000p11391}.
Information theoretic methods suggest that this type of ``intermittent noise" may permit information to be encoded in the spike patterns that single neurons produce over time~\cite{Reinagel:2000p11391,Tiesinga:2008p12846}. 

However, the impact of variability on network coding cannot be understood by extrapolating from single cells alone~\cite{Schneidman:2006p17360,Ecker2011a,Zoh+94,AverbeckLP06,abbott99,HuZSB:2013}.  Thus, to eventually understand how network chaos impacts coding, we need to capture the {\it multicell} spike train {variability} in chaotic networks -- and relate this to well-quantified measurements at the level of single cells. Direct, sampling-based approaches to this problem will fail, due to the combinatorial explosion of spike patterns that can occur in high-dimensional networks.  Another method is needed.


Studies of variability in recurrent networks typically address two distinct properties. On one hand, there is the question of spike-timing variability, often measured by binarized spike pattern entropy and usually studied for single cells or small cell groups~\cite{Schneidman:2006p17360,Reinagel:2000p11391,Strong:1998p15776}. On the other hand, recent theoretical work investigates the dynamical entropy production of entire networks, quantifying the state space expansion globally~\cite{Monteforte:2010p11768,Luccioli:2012p17675}. It is not clear how these two quantities are related. Here,  we extend the work of~\cite{Lajoie:2013p17297} to bridge this gap, leveraging random dynamical systems theory to develop a direct symbolic mapping between phase-space dynamics and binary spike pattern statistics.

 The result is a new bound for the variability of joint spike pattern distributions in large spiking networks that receive fluctuating input signals. This bound is in terms of spike-response noise entropy, an information-theoretic quantity that is directly related dynamical entropy production.  By verifying that the previous extensivity results of \cite{Monteforte:2010p11768,Luccioli:2012p17675} continue to hold in the presence of stimulus drive, we show how the bound applies to networks of all sizes.   

We then apply this bound to make two observations about the spike-pattern variability in chaotic networks.  The first is that the joint variability of spike responses across large networks is at least an order of magnitude lower than what would be extrapolated from measurements of spike-response entropy in single cells, despite noise correlations that are very low on average.   
Second, we show that the spike-response entropy of the network as a whole is strongly controlled by the tradeoff between the mean (i.e. DC) and higher-frequency components of the input signals. Entropy increases monotonically with the mean input strength by almost an order of magnitude, even as network firing rates remain constant.

\section{Network model}

To develop these results, we use large random networks of $N$ ``$\o$-neurons", as in~\cite{Monteforte:2010p11768, Lajoie:2013p17297}. The state of each cell is represented by a phase variable $\o_i(t)\in [0,1]$ where $0$ and $1$ are identified (ie. $S^1$) and a spike is said to occur when $\o_i=1\sim0$. This model has non-dimensionalized units but is equivalent to the Quadratic Integrate-and-Fire model via a smooth change of coordinates~\cite{Ermentrout:1996p10447}. In addition, the network receives a temporally structured input signal $I(t)$, as described below.

The dynamics of the $i^{\text{th}}$ cell in the network are given by the {\it random dynamical system} (RDS)
\beq
\label{net_model}
\begin{split}
d\o_i=&[F(\o_i)+Z(\o_i)\sum_{j=1}^Na_{ij}g(\o_j)+\frac{\e^2}{2}Z(\o_i)Z'(\o_i)]dt...\\
 &+Z(\o_i)\underbrace{[\n dt+\e dW_{i,t}]}_{I_i(t) dt}
\end{split}
\eeq
where $F(\o_i)=1+\cos(2\pi\o_i)$, $Z(\o_i)=1-\cos(2\pi\o_i)$ and 
\beqn
g(\o_j)=\left\{\begin{array}{cl}d \left( b^2 - \left[\left(\o_i+\frac{1}{2}\right)\text{mod}\,\, 1-\frac{1}{2}\right]^2 \right)^3 & \text{; }\o_i \in [-b,b] \\
0 & \text{; else}
\end{array}\right.
\eeqn
is a smooth coupling function with small support around $\o_j=1\sim0$, mimicking the rapid rise and fall of a synaptic current ($b=1/20$, $d=35/32$). The $\e^2$ term comes from an Ito correction~\cite{Lindner:2003p16585}. 

The network's input $I=\{I_i\}_{i=1}^N$, represented by the last term in~(\ref{net_model}), models a temporal stimulus. It is a collection of $N$ independent signals $I_i(t)=\n+\e dW_{i,t}/dt$ driving each neuron, where the $dW_{i,t}/dt$ are quenched realizations of white noise -- that is, scaled increments of the independent Wiener processes $W_{i,t}$. Note that $\n$ controls the network's ``excitability" and can take negative values~\cite{Ermentrout:1996p10447} while $\e\geq0$ controls the amplitude of input fluctuations.  Both parameters are constant across all cells.  We begin by investigating network~(\ref{net_model}) in the excitable regime with parameters $\n=-0.5$ and $\e=0.5$.
We emphasize that $I$ is a signal and not stochastic noise, and study the solutions of~(\ref{net_model}) arising from distinct initial conditions (IC) but receiving the same input $I$.  The model~\eqref{net_model} has been analyzed previously for uncoupled neurons~\cite{Ritt:2003p145,LinSY07a}, and for a series of gradually more complex networks in~\cite{LinSY07a,Lin:2009p16577,Lajoie:2013p17297}, cf.~\cite{Monteforte:2010p11768}.

 \begin{figure}[h!]
\begin{center}
\includegraphics{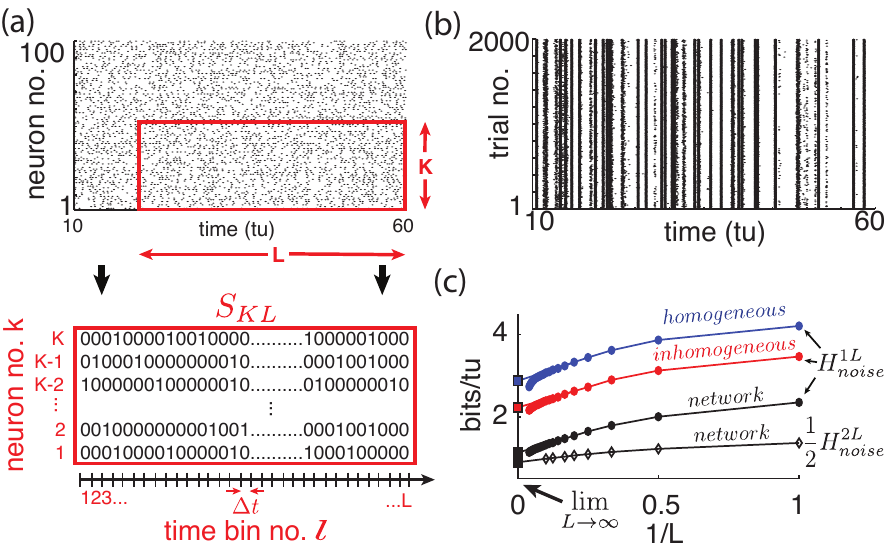}
\caption{(Color online) (A) Top: Raster plot of spike output for 100 randomly selected neurons on a single trial (dots are spikes). Bottom: Illustration of binary $S_{KL}$-word. (B) Raster plot of one randomly selected cell's spike output on 2000 trials where only network initial conditions change. (C) Single cell $H^{1L}_{noise}$ estimates for different choices of ``surrogate" noise (round markers); see text. From top to bottom: homogeneous poisson (blue), inhomogeneous poisson (red), network interactions (black). The bottom curve is a computation of $\frac{1}{2}H^{2L}_{noise}$ from a cell pair (diamond markers).  
Abscissa scale is $1/L$ to better visualize extrapolation of extensive regime to $L\to \infty$ (left square marker). 
For all panels: $\n=-0.5$, $\e=0.5$, $N=500$.}
\label{fig:general}
\end{center}
\end{figure}

We assign $20\%$ of the $N$ neurons to be inhibitory and $80\%$ to be excitatory, meaning that outgoing weights of neuron $j$ are either $a_{ij}\leq 0$ or $a_{ij}\geq 0$ respectively. The coupling matrix $A=\{a_{ij}\}_{i,j=1,...,N}$ is chosen randomly with mean {\it in-degree} $\k$ such that each neuron receives on average $\k$ incoming connections from independently chosen neurons, from each excitatory/inhibitory population. Here, $|a_{ij}|\sim \O(1/\sqrt\k)$ when non-zero, in accordance with classical {\it balanced state} coupling~\cite{Vreeswijk:1998p14451}. Throughout, we set $\k=20$ but find that as long as $\k \ll N$, our findings are qualitatively robust to the choice of $\k$. 

Two consequences of this connectivity will be important below. First, as the in-degree $\k$ is the same for all neurons, the spiking statistics of single cells are fairly stereotypical across the network. 
This is evident in the spike rasters of Figure~\ref{fig:general} (a). Second, the magnitude of inputs to single cells remains similar as network size $N$ grows, because $\k$ is fixed.

\section{Spike-response noise entropy and direct estimates}
To quantify spike pattern variability, we treat spike trains as binary time series.  We discretize time in bins of width $\D t$ small enough so that for a given cell, each bin contains at most a single spike. Throughout, we use time bins of width $\D t=0.05$; we found that moderately different resolutions did not significantly affect our results. Let us define finite binary words for $K$ neurons over $L$ time bins starting at time $t_l=l\D t$ for some integer $l$: $S_{KL}(t_l)=\{S_l^{k},...,S_{l+L-1}^{k}\}_{k=k_1,...,k_K}$ with $S_{j}^k \in \{0,1\}$ (see Figure~\ref{fig:general} (a)).

The variability of the evoked spike response $S_{KL}(t_l)$ is captured by the {\it spike-response noise entropy}
\beq
\label{time_H}
H_{noise}^{KL}(I,t_l)=\frac{-1}{L\D t}\sum_{S_{KL}}P(S_{KL}(t_l)|I)\log_2P(S_{KL}(t_l)|I)
\eeq
where $P(S_{KL}(t_l)|I)$ denotes probability of observing word $S_{KL}(t_l)$ conditioned on input $I$, given a random initial state of the network. This quantity may also be referred to as conditional response entropy. It is normalized to have units of bits per time-unit ($bits/tu$), as opposed to bits per time-bin, and thus represents an {\it entropy rate} in continuous time. 
Since the inputs $I$ and network dynamics are statistically stationary processes~\cite{Lajoie:2013p17297}, it follows that the expected noise entropy rate of $KL$ words conditioned on any $I$ from the same input distribution --- controlled by the parameters $\n$ and $\e$ --- can be obtained from a long time average on any single $I^*$ (see eg.~\cite{Strong:1998p15776,spikesbook}):
\beq
\label{H_KL}
H_{noise}^{KL}=\int_IP(I)H_{noise}^{KL}(I,t_l)=\lim_{T \to \infty}\frac{1}{T}\sum_{l=0}^{T-1}H_{noise}^{KL}(I^*,t_l).
\eeq

As demonstrated in~\cite{Strong:1998p15776} and reviewed below,~(\ref{H_KL}) can be used to estimate the true entropy rate of $K$-neuron groups considered when $L \to \infty$.  As we will see this is only practical for small $K$ --- we will need other tools to understand this quantity for entire networks ($K=N$). Nevertheless, we begin by applying a direct sampling approach.


To estimate the probability terms in~(\ref{time_H}),
we simulate network~(\ref{net_model}) in response to a randomly chosen, quenched $I(t)$ for $10,000$ time units and $2000$ ``trials", distinguished by different ICs. Here, we wish to choose ICs from a distribution that best describes random network states, while being agnostic about its past.
As discussed in~\cite{Lajoie:2013p17297}, we assume that system~(\ref{net_model}) possesses an ergodic stationary probability measure $\mu(\o)$, which is the steady state solution of the Fokker-Planck equation associated with~(\ref{net_model}). Thus, $\mu$ is the probability measure describing how likely we are to find the network in a particular state at any moment in time, given any input $I$ with identical statistics. 
We emphasize that $\mu$ serves only as an initial distribution, and that ensembles of ``trial" trajectories as described above will have a very different distribution, as they are conditioned on a {\it fixed} input $I(t)$.  (See~\cite{LinSY07a,Lajoie:2013p17297,Lin:2009p16577} for more details about this distinction).


%

To sample from $\mu$, we first select seed ICs uniformly over the state space, and evolve each of these for a ``burn" period of $50$ time units, for which different inputs are presented. The resulting endpoints of these trajectories represent a new IC ensemble that approximates $\mu$. From then on, all ICs are integrated using the same input $I(t)$ and we use this solution ensemble to study variability of spike-responses. 

From these simulated network trajectories, we extract the binary spike output of neurons across many trials (see Figure~\ref{fig:general} (b) for a single neuron example).  
Normalized cross-trial counts of $S_{KL}$ words in consecutive, non-overlapping $L$-windows serve as estimates of the probabilities $P(S_{KL}(t_l)|I)$ in equation~(\ref{time_H}).

\section{Single-cell variability}


We begin by computing noise entropy in the spike responses of single cells in the network.  Using the estimation techniques described above, we compare the effect of chaos to that of commonly used independent noise models on noise entropy. This complements similar analysis in~\cite{Lajoie:2013p17297}, which used a different metric of spike reliability from trial to trial.

We begin by randomly selecting a cell in our network and extract its binary spike output across many simulated trials (see Figure~\ref{fig:general} (b)). Using this data, we estimate $H_{noise}^{1L}$ for word lengths up to $L=20$ and plot the results in Figure~\ref{fig:general} (c) as a function of $1/L$. A system with finite autocorrelation timescales is expected to produce entropy rates that behave extensively as $L$ becomes sufficiently large. This is readily apparent in the linear decreasing trend in $H_{noise}^{1L}$ as $L$ grows, until a point where the estimate quickly drops due to insufficient sampling. Following~\cite{Strong:1998p15776}, we use the point of least fractional change in slope to extrapolate this extensive trend and obtain an estimate for $\lim_{L\to \infty}H_{noise}^{1L}$ (intersection with ordinates in Figure~\ref{fig:general} (c)).  

 Our estimate of $\lim_{L\to \infty}H_{noise}^{1L}$ is $1.12$ $bits/tu$.  We note that a ``purely random", homogeneous poisson spike train with the same firing rate ($0.8$ $spikes/tu$) would have noise entropy $H_{noise}^{1L}$ of $3.67$ $bits/tu$.  Thus, while chaotic dynamics produce variable spiking in single cells, the resulting noise entropy is much less than that of a totally random response, a fact also evident from the spike rasters in Figure~\ref{fig:general} (b).

Part of the reason for this difference is simply the presence of the stimulus; inputs from other cells in the chaotic networks also play a role.  To isolate the network effect, we repeat the sampling process above by simulating our chosen cell in isolation, keeping the input $I_i$ intact but replacing the incoming spike trains it receives from upstream cells by two surrogate ensembles meant to isolate distinct statistical features of network activity. (i) {\it Homogeneous poisson} surrogates: independent, poisson distributed spike trains with rate matching the mean firing rate of corresponding upstream cells. (ii) {\it Inhomogeneous poisson} surrogates: produced by independently drawing a binary random variable in each $\D t$-bin, according to the time-dependent probability given by the normalized spike count of the corresponding network train across all original trials. For each new simulated trial, we draw independent surrogates. 
Figure~\ref{fig:general} (c) shows a $66\%$ increase in noise entropy rate for the homogeneous surrogates, and about $30\%$ for the inhomogeneous case. 

Overall, we have shown that single, stimulus-driven cells in chaotic networks produce spike-response entropy significantly lower than that expected for single, stimulus-driven cells receiving poisson background inputs, as in many statistical models.    
We next seek to characterize spike entropy in the joint responses of multiple cells.  

\section{Multi-cell variability}

Our network is connected --- albeit sparsely ($\k \ll N$) --- and it is not clear in advance how coupling interactions will impact the entropy rate of groups of cells. As a first step, we repeat the noise entropy estimate described above for a randomly selected pair of connected cells up to $L=10$, and extrapolate $\lim_{L\to \infty}H_{noise}^{2L}$ from this data.  The black lines in Figure~\ref{fig:general} (c) show $H_{noise}^{2L}/2$, normalized to units of bits per time-unit per neuron for comparison with $H_{noise}^{1L}$. Due to combinatorial explosion of possible spike patterns as more neurons are considered, we were unable to compute such estimates for $K$ greater than 2. Nevertheless, it appears from the $K=2$ case shown that interactions between neurons conspire to lower response noise entropy per neuron, if only by a small margin.

However, this margin could easily be missed. For a given neuron pair $(i,j)$, consider the difference between the sum of independent cell entropy rates and their joint pair rate: $\d_{ij}=\lim_{L\to\infty}[H_{noise}^{1L}(i)+H_{noise}^{1L}(j)-H_{noise}^{2L}(i,j)]$. From 45 random pairs of neurons, we obtain the average $\langle \d_{ij} \rangle=0.012$ $bits/tu$. This implies a relative difference of the order of $\O(10^{-2})$ when estimating the entropy rate of pairs of cells using their marginal, single-cell response distributions. We will see later these small differences compound significantly when considering the network as a whole (cf.~\cite{Schneidman:2006p17360}).



\begin{figure}[h!]
\begin{center}
\includegraphics{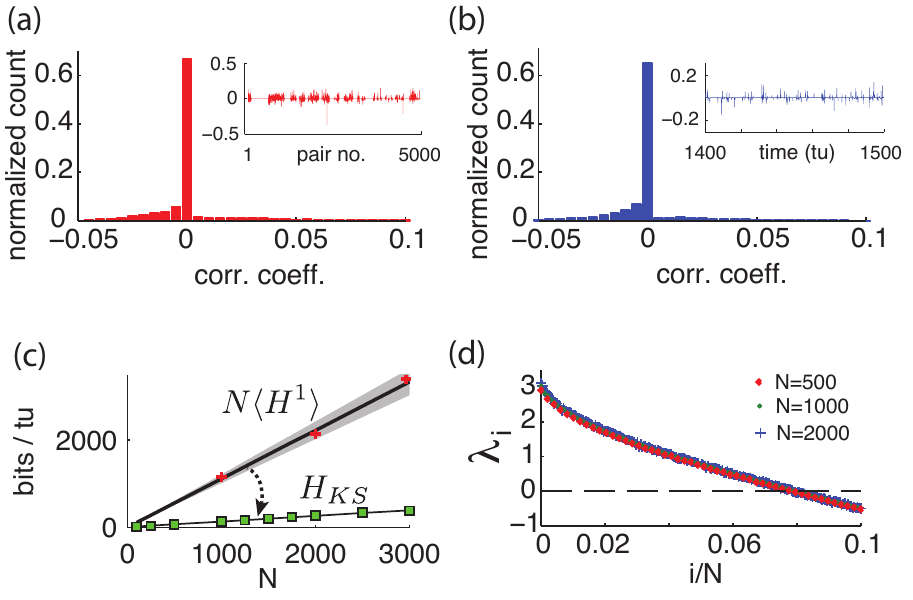}
\caption{(color online) (A) Typical histogram of noise correlation coefficient $c_{ij}(t_l)$ between all neuron pairs for a fixed time. Inset shows $c_{ij}(t_l)$ for the first $5000$ pairs. (B) Histogram of noise correlation coefficient $c_{ij}(t_l)$ between two connected cells across $10,000\,tu$. Inset shows $c_{ij}(t_l)$ for $100$ $tu$. (C) Network-wide noise entropy estimates in $bits/tu$ as a function of $N$. Slope $\langle H^1\rangle$ averaged over 20 random cells in a network with $N=500$. Shaded area shows two standard errors of the mean. Markers show direct samples of single cells for various network sizes. $H_{KS}$: square markers shows estimates from Lyapunov spectra for a range of $N$; black line is a linear fit. (D) Plot of first $10\%$ of Lyap spectrum for $N=500$, $1000$ and $2000$ .For all panels: $\n=-0.5$, $\e=0.5$.}
\label{fig:KS}
\end{center}
\end{figure}

To quantify the extent of these interactions over space and time, we compute the Pearson correlation coefficient $c_{ij}(t_l)$ between the spiking probability of two cells $i$ and $j$ in time bin $t_l$.  That is, we measure the cells' instantaneous {\it noise correlation}. 
Figure~\ref{fig:KS} (a) shows a typical histogram of $c_{ij}(t_l)$ across all neuron pairs of a network with $N=500$ for a fixed $t_l$, where pairs with zero spiking probability were discarded.  We can see that at a fixed moment, correlations are weak and most cells are uncorrelated.  Moreover, these correlations are not static:  a high correlation between two cells in one time bin does not guarantee that they will be correlated in another. This is illustrated by Figure~\ref{fig:KS} (b), showing a histogram of $c_{ij}(t_l)$ across $10000$ time-units between two randomly chosen connected cells. 

We emphasize that this weak and highly dynamic correlation structure might easily be dismissed as negligible experimentally.   If one would choose a single pair of cells and measure the temporal average of $c_{ij}(t_l)$ over 500 time units, one obtains an average of the order of $10^{-5}$ (over 4950 cell pairs tested), and standard deviation of the order of $10^{-2}$ (across the 4950 cell pairs.)  In other words, each individual cell pair appears to be almost completely uncorrelated -- at least on average.
Below, we will show that the weak, transient dependencies that are in fact present among neurons nevertheless have a very strong impact on network-wide noise entropy.

To summarize, measures of entropy and correlations indicate that there are noticeable but weak dependencies in the spiking activity of connected pairs of cells.  Scaling up from such dependencies to accurately describe the joint activity of an entire network is a notoriously difficult problem.  We take an approach based on RDS in what follows.  This approach will quantify the entropy $H^{NL}_{noise}$ of the network as whole, as networks size $N$ grows.  

\section{A benchmark for network entropy}
To benchmark $H^{NL}_{noise}$, we first describe the joint network entropy that would be naively predicted by direct extrapolation from single cells. In other words, this is the estimate one would obtain by ignoring statistical interactions between neurons. Notice that unlike cell pairs, spiking statistics of single neurons are expected to be unchanged by network size $N$ with fixed in-degree $\k$.  Moreover, the entropy of a multivariate distribution is always greater or equal to the sum of the marginal distributions' entropies.
If $\langle H^1 \rangle$ denotes the average of $\lim_{L\to\infty}H^{1L}_{noise}$ over all neurons, then it follows that $N\langle H^1\rangle  \geq \lim_{L\to\infty}H^{NL}_{noise}$. 
Figure~\ref{fig:KS} (c) shows this estimate as a function of $N$. The slope $\langle H^1\rangle$ was sampled over $20$ neurons in a $N=500$ network using the same extrapolation technique as in Figure~\ref{fig:general} (c). We verified by spot checks that single cell activity in networks of different sizes agree with this extrapolation (see markers in Figure~\ref{fig:KS} (c)). Next, we leverage dynamical properties of our network to estimate how much reduction in entropy can be expected from the joint activity of entire networks in comparison to this naive extensive bound.

\bigskip

\section{Dynamical entropy production}


In what follows, we use symbolic dynamics to map between the phase space of our network and the set of binary spike trains.
Consider trajectories $\o(t)=(\o_1(t),...,\o_N(t))$ of model~(\ref{net_model}), evolving on the $N$-dimensional torus $\T^N$. Recall that a spike from cell $i$ occurs when $\o_i(t)=1$, and will lead to $S_l^i=1$ in the corresponding time bin. Notice that the phase response curve $Z(\o_i)$ modulates the effect of any input on neuron $i$ -- whether that input comes from the signal $I(t) $ or from network activity -- and that it vanishes at $\o_i(t)=1$. This implies that a neuron becomes insensitive to any inputs when it is about to spike. Indeed, the Taylor expansion of neuron $i$'s dynamics about $\o_i=1$ is constant up to quadratic order: $d\o_i=[2+\O((\o_i-1)^2)]dt+\O((\o_i-1)^2)dW_{i,t}$. Based on this observation we make the approximation that for $\D t$ small enough, neuron $i$ spikes in the time bin $[t,t+ \D t]$ if and only if $\o_i(t) \in [1-2\D t, 1)$ (see Appendix for verification). 

Thus equipped, consider the partition of $\T^N$: $\G^*=\{\g_0,\g_1\}^N$, built of Cartesian products of intervals $\g_0=[0,1-2 \D t)$ and $\g_1=[1-2\D t,1)$ across all $\o_i$'s. At any time $t_l=l\D t$, the $\G^*$-address of $\o(t_l)$ determines the binarized spiking state of the network in time bin $[t_l, t_l+\D t]$: $\o_i(t_l)\in \g_0 \Rightarrow S^i_l=0$ and $\o_i(t_l)\in \g_1 \Rightarrow S^i_l=1$. In order to describe $L$-long spike trains in terms of $\G^*$-addresses, we must understand how solutions $\o(t)$ evolve with respect to $\G^*$. 
To this end, consider the discretized dynamics given by the transition maps $\Phi_{t;I}$ that send $\T^N$ onto itself according to the flow of~(\ref{net_model}) from $t$ to $t+\D t$. If $\o(t)$ is a solution of~(\ref{net_model}), then $\Phi_{t;I}(\o(t))=\o(t+\D t)$ where $\D t$ refers to the resolution of our binary spike trains $S_{NL}$. Note that the maps $\Phi_{t;I}$ depend on both $t$ and $I$, are generally smooth with smooth inverses (diffeomorphisms)~\cite{kunita}, and together form a discrete RDS. For detailed geometric properties of the RDS defined by system~(\ref{net_model}), we refer the reader to~\cite{Lajoie:2013p17297}.

For what follows, it is convenient to reverse time and study spike trains and trajectories starting in the distant past leading up to $t=0$. This representation is statistically equivalent to forward time since our networks have statistically stationary dynamics~\cite{Lajoie:2013p17297}.
Consider now the $l$-step inverse map: $\Phi_{0;I}^{-l}$. For any set $A$ in the partition $\G^*$, its pre-image $\Phi_{0;I}^{-l}(A)$ refers to all points in $\T^N$ at time $-l\D t$ that will be mapped to $A$, and consequently have the same spiking state at $t=0$. Similarly, if both $A_0$ and $A_1$ are sets in $\G^*$, the intersection $\Phi_{0;I}^{-l}(A_0) \bigcap \Phi_{0;I}^{-l+1}(A_1)$ describes all points that will be mapped to $A_1$ at $t=-\D t$ and $A_0$ at $t=0$. It follows that any subset of the form $B=\bigcap_{l=0}^{L} \Phi_{0;I}^{-l}(A_s)$ where $A_s\in\G^*$ captures all past network states at time $t=(-L)\D t$ leading to identical spiking sequences $\{S^i_{-L},...,S^i_{-1},S^i_0\}_{i=1,..,N}$, when the same $I$ is presented. Moreover, it is easy to show that the collections of all possible sets constructed as $B$, named the {\it join} of pre-images of $\G^*$ denoted $\vee_{l=0}^{L}\Phi_{0;I}^{-l}\G^*$, is itself a partition of $\T^N$.

It follows that this new partition offers a one-to-one correspondence between its member sets and the space of all $S_{NL}$ spike trains. Note that many sets in this partition will be empty since not all spike sequences are accessible by the network. In fact, the number of non-empty sets remaining in $\vee_{l=0}^{L-1}\Phi_{0;I}^{-l}\G^*$ as $L\to\infty$ represents the number of allowed infinite spike sequences. Furthermore, for a given $S_{NL}$ and its associated set $B(S_{NL}) \in \vee_{l=0}^{L-1}\Phi_{0;I}^{-l}\G^*$, the probability of observing spike pattern $S_{NL}$ can be stated as an initial state probability in the distant past: 
$P(S_{NL}|I)=P(\o(-L\D t)\in B(S_{NL}))$. 

As discussed above and in~\cite{Lajoie:2013p17297}, we assume that our RDS possesses an ergodic stationary probability measure $\mu$. Recall that we assume random ICs forming our distinct trials are drawn from $\mu$. It follows that $\lim_{L\to \infty} P(S_{NL}|I)=\mu(B(S_{NL}))$. Thus, if we let
\beq
\label{part}
h_\mu(\Phi_{t;I},\G^*)=\lim_{L\to \infty} -\frac{1}{L}\sum_{B \in \vee_{l=0}^L\Phi_{0;I}^{-l}\G^*}\mu(B)\ln\mu(B),
\eeq
it follows that 
\beq
\label{spike_part}
\lim_{L\to\infty}H_{noise}^{NL}=  \frac{\D t}{\ln2}h_\mu(\Phi_{t;I},\G^*).
\eeq

For any dynamical system, the expression~(\ref{part}) measures the amount of uncertainty produced by chaotic dynamics if we can only observe the system with the precision given by the partition $\G^*$. This concept is generalized by the {\it Kolmogorov-Sinai entropy} $h_\mu$, also called {\it dynamical} or {\it metric} entropy~\cite{entropy, Ruelle:booklet}, defined by
\beq
\label{KS}
h_\mu = \sup_\G h_\mu(\Phi_{t;I},\G)
\eeq
where the supremum is taken over all finite partitions $\G$. This quantity is related to the Lyapunov spectrum $\l_1\geq\l_2\geq...\geq\l_N$ of a dynamical system which measures rates of exponential divergence or convergence between trajectories. Lyapunov exponents $\l_i$ are expected to be well defined for our RDS in the sense that they generally on system parameters on system parameters such as coupling strength and the mean and variance of inputs, but not on specific realizations of the inputs $I(t)$~\cite{kifer}. The authors of~\cite{Led+88} showed that although the join of a partition $\G$ depends on $I$, $h_\mu$ does not and that under some ergodicity assumptions, the following entropy formula holds:
\beq
\label{h_formula}
h_\mu = \sum_{\l_i>0} \l_i.
\eeq
If $\l_i$ are the Lyapunov exponents of the original system~(\ref{net_model}) computed over time-units instead of $\D t$ time-steps, we get from~(\ref{part}),~(\ref{spike_part}),~\eqref{KS} and~(\ref{h_formula}) the following upper bound for noise entropy rate :
\beq
\label{KS_bound}
H_{KS}\equiv\frac{1}{\ln2}\sum_{\l_i>0} \l_i\geq \lim_{L\to\infty}H_{noise}^{NL} 
\eeq
which has units of bits per time-unit.

To evaluate this bound, we numerically compute the exponents $\l_i$ of system~(\ref{net_model}) and find that, as originally observed in~\cite{Monteforte:2010p11768,Luccioli:2012p17675} for autonomous networks, our driven system has a size invariant Lyapunov spectrum (see Figure~\ref{fig:KS} (d)), which is insensitive to particular choices of random coupling matrix $A$ (see Appendix for details). This leads to a spatially extensive behaviour of the bound $H_{KS}$, as shown in figure~\ref{fig:KS} (c). 

Intriguingly, $H_{KS}$ is much smaller than estimates from $\langle H^1\rangle$.  This reveals a central result for our driven chaotic networks: {\it joint spike patterns are (at least) an order of magnitude less variable than what would be predicted by observing the spike train statistics of single cells, despite averaged noise correlations across neurons that are very low}. 

\section{Noise entropy production as a function of input statistics}
\begin{figure}[h!]
\begin{center}
\includegraphics{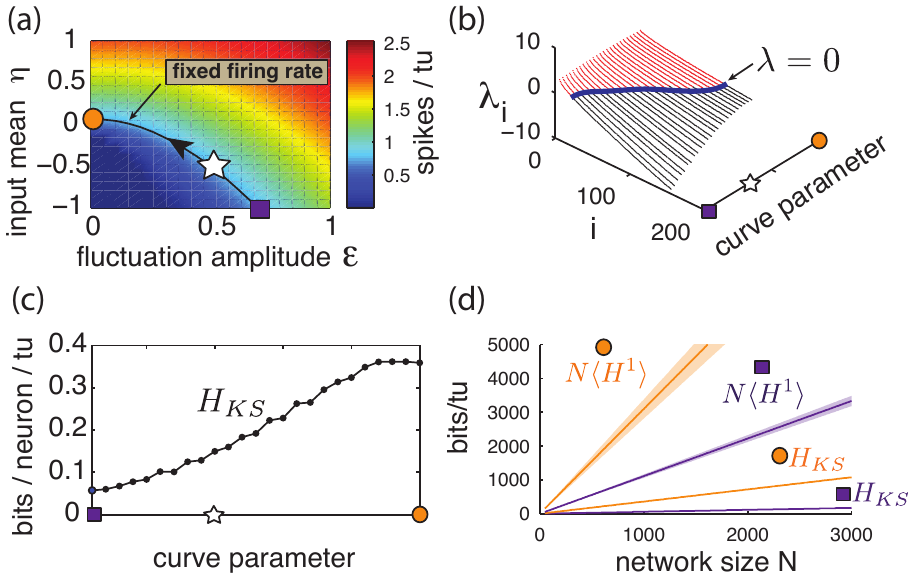}
\caption{(color online) (A) Heat map of excitatory population mean firing rate for a range of input amplitude $\e$ and input mean $\n$. Line is the contour curve for fixed firing rate of $0.820$ $spikes/tu$ $\pm$ $0.003$, parametrized by numerical interpolation. Arrow shows direction of parametrization. Markers: square:  $\n=-1$, $\e=0.69$, star: $\n=-0.5$, $\e=0.5$, circle: $\n=0.07$, $\e=0$. (B) Lyapunov spectra along contour curve from (a). (C) $H_{KS}$ bounds evaluated along contour curve from (a). (D) Network noise entropy bounds $N\langle H^1\rangle$ and $H_{KS}$ for square and circle marker parameters in (a). Slope $\langle H^1\rangle$ averaged over 20 random cells. Shaded area shows two standard errors of the mean. Both $\langle H^1\rangle$ and $H_{KS}$ extrapolated  from a network with $N=500$, as are quantities from all other panels.}
\label{fig:regime}
\end{center}
\end{figure}

Previous studies showed that the level of sensitivity emerging from chaotic network dynamics can be controlled by carefully chosen inputs (see~\cite{Rajan:2010p7924,PhysRevLett.69.3717} for different contexts). We verify if this is the case for our network. We first identify a range of input statistics --- the mean $\n$ and fluctuation amplitude $\e$ --- that are comparable in that they all produce the same firing rate as for the ``standard" parameter set used above ($\n=-0.5$, $\e=0.5$).  These parameters lie along the level curve in Figure~\ref{fig:regime} (a).  Note that the curve is parametrized so that $\n$ grows while $\e$ decreases; thus, as we travel along it, we gradually shift the dynamics from the excitable, fluctuation-driven regime ($\n<0$) to an oscillatory, mean-driven one ($\n>0$). In particular, the last point evaluated corresponds to a purely autonomous regime ($\e=0$) where the input $I$ has no fluctuating component.

Figure~\ref{fig:regime} (b) shows the first $200$ Lyapunov exponents of a network with $N=500$ along this level curve, and panel (c) gives the corresponding $H_{KS}$ values.  A clear trend emerges: $H_{KS}$ increases monotonically as the system transitions from fluctuation- to mean-driven regimes, by almost an order of magnitude.  Moreover, Figure~\ref{fig:regime} (d) shows that, for the two extremes of the level curve, network noise entropy continues to be much smaller than that predicted from single cells, and that single-cell noise entropy appears to follow the same trends as $H_{KS}$.  We conclude that {\it spike pattern variability emerging from chaos is not  a fixed property of a network, but can be strongly modulated by the mean and variance of network inputs.}

\section{Discussion}




Biological neural networks may operate in a chaotic regime, with irregular activity driven by a balance of fluctuating excitatory and inhibitory interactions.   This network chaos is under vigorous study, fueled in part by possible roles for chaos in generating ``target" spatiotemporal patterns~\cite{Sussillo:2030p4602} and in enabling useful temporal processing of inputs~\cite{Buonomano:2009p17775,Laje:2013p17820}. Here, we address a complementary question -- how much variability (or ``noise") will chaotic dynamics add to network responses?


We compute bounds on network spike-response entropy that give novel answers. In particular, we show that the noise entropy of multi-cell spike responses is at least an order of magnitude lower that would be naively extrapolated from from single-cell measurements, under the assumption that spike variability is independent from cell to cell. The ÒdirectionÓ of the comparison between noise entropy of single cell and multi-cell spike responses agrees with intuition provided by the shape of the Lyapunov spectrum, which indicates time-dependent chaotic attractors of lower dimension than phase space. Thus, the phase space dynamics of each neuron are not independent. What we quantify explicitly is the order-of-magnitude size of the effect, as it is manifested in the binary {\it spiking} outputs of the system --- a fact which might seem especially striking given that pairs of spike trains appear to be very weakly correlated on average.

If one considers the level of noise entropy as an indicator of potential information contained in spike patterns, we show that balanced networks may be able to encode inputs stimuli using spike timing if these inputs contain strong enough temporal structure. 
This mechanism takes root in the complex noise-interactions that chaos induces between neurons.
The extensive nature of this phenomenon suggests that this mechanism is scalable with network size. Moreover, the strong dependence of entropy on the input signal's mean and variance indicate that a network can operate in different ``regimes" modulating the repeatability of spike patterns. This is in addition to known advantages of balanced networks, such as efficiently tracking changes in common, mean inputs with firing rates~\cite{Vreeswijk:1998p14451} --- which may encode coarser statistics about inputs at the population level.

To formalize these notions, future work could seek to compute the mutual information between an input ensemble and a system's response. In order to estimate this quantity, one needs to compute the {\it total entropy}~\cite{spikesbook} of spike patterns, which captures how many distinct spike outputs can be produced by the network, for any input $I$. This quantity can be thought of as noise entropy marginalized over the set of possible inputs, and therefore depends on both network connectivity and single neuron attributes.

Finally, we expect that the $H_{KS}$ bound can be adapted to other neuron models provided a state space partition linking dynamics to spike patterns can be derived. This could prove to be a powerful tool to enquire about potential encoding schemes as a function of many network attributes such as spike-generating dynamics, connectivity, learning rules and input correlations. 



 \section{Acknowledgments}
The authors thank Fred Wolf, Yu Hu and Kevin K. Lin for helpful insights. We also thank two anonymous reviewers for comments and suggestions that improved the manuscript.  This work was supported in part by an NSERC graduate scholarship, an NIH Training Grant from University of Washington's Center for Computational Neuroscience, the Burroughs Wellcome Fund Scientific Interfaces, the NSF under grant DMS CAREER-1056125 and NSERC Discovery and CIHR operating grants. Numerical simulations were performed on NSF's XSEDE supercomputing platform.

\appendix
\section{APPENDIX}


\section{Numerical simulations}
 Throughout the main text, we use data from numerical simulations of the network model described by~(\ref{net_model}).
%
All simulations were implemented using a standard Euleur-Maruyama solver with time-steps of $0.005$ time-units. We found that using smaller time-steps did not alter our results. The solver was developed using the Python/Cython programming language using the Mersenne Twister random number generator and post-processing (spike binning and empirical noise entropy estimates) was carried out in MATLAB. Large simulations were performed on the NSF XSEDE {\it Science Gateways} supercomputing platform.

\section{Lyapunov spectrum estimates}

\begin{figure}[h!]
\begin{center}
\includegraphics{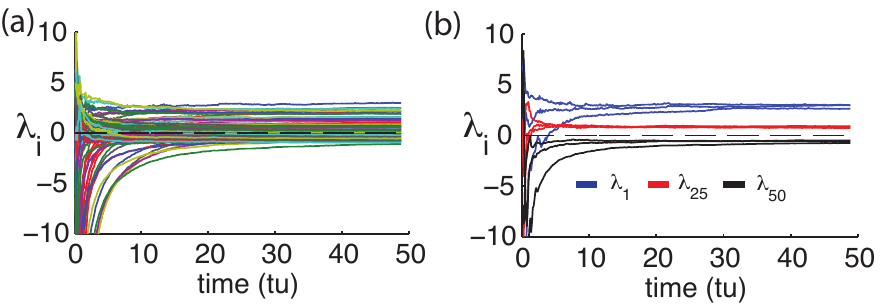}
\caption{Estimates of Lyapunov exponents for the initial $50$ out of $5000$ time-units, showing convergence.  (a) Estimates of the first $60$ Lyapunov exponents (out of $500$) for a given network. (b) Three distinct estimates for $\l_1$, $\l_{25}$ and $\l_{50}$ where network IC, $I$ and coupling matrix $A$ are selected differently and at random. For both panels, $N=500$, $\e=0.5$, $\n=-0.5$.}
\label{fig:lyap}
\end{center}
\end{figure}

Although the Lyapunov exponents $\l_1\geq\l_2\geq...\geq\l_N$ of~(\ref{net_model}) do not depend on a particular choice of $I$ or initial conditions (IC), computing them analytically is a very hard, if not an impossible, problem. Therefore, we use numerical estimates. While numerically integrating a solution of~(\ref{net_model}) above, we simultaneously evolve the linear variational equation
\beq
\label{var_eq}
\dot{M}=J(t)M
\eeq
where $J(t)$ is the Jacobian of~(\ref{net_model}) evaluated along the simulated trajectory. Here, $M$ is a $N$ by $N$ matrix where $M(0)$ is the identity. $M(t)$ is orthonormalized at each time-step and the growth factors of each orthogonal vector obtained from the process are extracted to build estimates that converge toward the $\l_i$'s, as described in~\cite{Geist:1990p12843}. This process was repeated for ten random choices of the input $I$ and the initial states; trajectories were integrated for $5000$ time-units. We verified that all reported $\l_i$'s have a standard error less than $0.002$ using the method of batched means~\cite{Asmussen2007} (batch size of $100$ time-units). Figure~\ref{fig:lyap} (a) shows converging estimates of the first $60$ Lyapunov exponents over the initial $50$ time-units.

In addition, we find that distinct realizations of connectivity matrix $A=\{a_{ij}\}$ did not significantly affect the Lyapunov exponent estimates --- and hence the sum of all positive ones leading to the Kolmogorov-Sinai entropy $h_\mu$. To illustrate this, Figure~\ref{fig:lyap} (b) shows estimates of three $\l_i$'s for three distinct systems, where input choice $I$, IC and $A$ are all different.

\section{Relationship between state space partitioning and spiking patterns}
\label{piecewise}
The derivation of the $H_{KS}$ bound relies on the simple assumption that neuron $i$ will spike within $\D t$ time-units if and only if $\o_i(t) \in \g_1=[1-2\D t,1]$. We found that for simulated trajectories of $1000$ time-units from network~(\ref{net_model}), only about $0.01\%$ of all spikes violated the spiking assumption for $\D t=0.05$. This number dropped to zero for $\D t=0.01$.  
Such values are evidence that errors in relating spike train entropy estimates to entropy production in state space will be slight.    
As an additional check, we next compare the spiking statistics and entropy estimates for the main model~(\ref{net_model}) with those for an analogous dynamical system, for which our partition-based spiking assumption holds exactly, by design.

Consider the {\it piecewise} model analogous to system~(\ref{net_model}):
\beq
\label{piece}
\begin{split}
d\o_i=&[\tilde{F}(\o_i)+\tilde{Z}(\o_i)\sum_{j=1}^Na_{ij}g(\o_j)+\frac{\e^2}{2}\tilde{Z}(\o_i)\tilde{Z}'(\o_i)]dt...\\
&+\tilde{Z}(\o_i)\underbrace{[\n dt+\e dW_{i,t}]}_{I_i(t)}
\end{split}
\eeq
in which we replace the functions $F$ and $Z$ by the following piecewise-defined terms:
\beqn
\begin{split}
\tilde{F}(\o_i)&=\left\{\begin{array}{cl}1+\cos(2\pi\o_i) & \text{; }\o_i \in [0,1-2\D t) \\
2 & \text{; }\o_i\in[1-2\D t,1)
\end{array}\right.\\
\tilde{Z}(\o_i)&=\left\{\begin{array}{cl}1-\cos(2\pi\o_i) & \text{; }\o_i \in [0,1-2\D t) \\
0 & \text{; }\o_i\in[1-2\D t,1).
\end{array}\right.
\end{split}
\eeqn
It is easy to see that the partition-based spiking assumption holds exactly for the network defined by~(\ref{piece}). However, notice that for $\D t>0$, both $\tilde{F}$ and $\tilde{Z}$ are discontinuous functions of $S^1$ and that as a result, the Jacobian of~(\ref{piece}) is ill-defined. Nevertheless, for practical purposes, we can simulate system~(\ref{piece}) and approximate its Lyapunov spectrum, since there is only one discontinuity point per neuron and the probability of a finite-duration, discretized trajectory landing on such points is nil.
 \begin{figure}[h!]
\begin{center}
\includegraphics{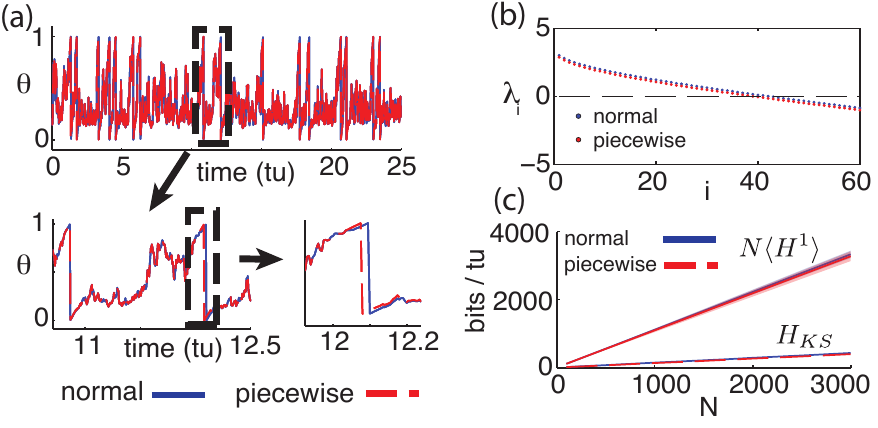}
\caption{(a) Comparison of trajectories for single cells, for models~(\ref{net_model}) and~(\ref{piece}); initial conditions and inputs are fixed. (b) First $60$ Lyapunov exponents of models~(\ref{net_model}) and~(\ref{piece}). (c) Empirical noise entropy bounds $NH^1$ and $H_{KS}$ for models~(\ref{net_model}) and~(\ref{piece}). For all panels, $\n=-0.5$, $\e=0.5$, $\D t=0.05$. For panels (b) and (c), $N=500$, $\k=20$.}
\label{fig:piece}
\end{center}
\end{figure}

The purpose of model~(\ref{piece}) is to assess the differences arising between the dynamics of our full (``normal") model, given by Eqn.~(\ref{net_model}), and the alternate (``piecewise") model above for which the spiking assumption is exact. We fix $\D t=0.05$ as in the main text and begin by comparing single cell dynamics for the ``normal" and ``piecewise" models. Figure~\ref{fig:piece} shows a simulated single cell trajectory from each model, with identical input $I_i$ and identical incoming spike trains (extracted from a separate network simulation). This setup mimics the activity a single cell would receive when embedded in a network. Notice that apart from small discrepancies that sometimes arise between spike times, the two trajectories agree almost perfectly. When differences do arise, they are quite small. From a simulation yielding about $3000$ spikes from both models, most corresponding spikes from the normal and piecewise models were indistinguishably close, down to the numerical solver's time-step. The maximal difference was about $0.02$ time-units, smaller than a $\D t$ time-bin. 

 
 
 Figure~\ref{fig:piece} (b) shows the first $60$ Lyapunov exponents of a network with size $N=500$, simulated with both the normal~(\ref{net_model}) and piecewise~(\ref{piece}) models. Since Lyapunov exponents depend on the Jacobian of a system, we expected the piecewise model to yield smaller exponents:  its derivative is zero on the intervals $[1-2\D t,1)$. Nevertheless, this discrepancy is minimal and amounts to a difference of about $0.002$ bits per neuron per time-unit in the slope of the $H_{KS}$ estimates shown in Figure~\ref{fig:piece} (c). Finally, we empirically estimate the noise entropy bound $\langle H^1 \rangle$, as described in the main text, for the piecewise model~(\ref{piece}). Its value differed from the normal model estimate by about $0.01$ bits per neuron per time-unit, well below the standard error of the mean of estimates from both models, as can be seen in Figure~\ref{fig:piece} (c).

In light of these tests, we are confident that the main result of the paper --- a computable bound on spike-train noise entropy that is much lower than what would be extrapolated from single cells --- is a robust phenomenon for networks of the type modeled by~(\ref{net_model}), rather than a consequence of a (seemingly tiny) approximation error.

%

\end{document}